\documentclass[a4,useAMS,usenatbib,usegraphicx]{mn2e}
\def\apss{\ref@jnl{Ap\&SS}} 
\def\aapr{\ref@jnl{A\&A~Rev.}}		

\usepackage{amsmath}
\include{epsf}
\usepackage{graphics}
\usepackage{epsf}
\usepackage{rotate}
\epsfclipon

\title[Harassment on infalling late-type dwarfs]{How effective is harassment on infalling late-type dwarfs?}
\author[R.Smith et al.]{R.Smith$^{1,2}$\thanks{E-mail:
rsmith@astro-udec.cl}, J.I.Davies${^1}$, A.H.Nelson${^1}$\\
$^{1}$School of Physics \& Astronomy, University of Wales, Cardiff, U.K\\
\noindent
$^{2}$Departamento de Astronomia, Universidad de Concepcion, Casila 160-C, Concepcion, Chile}
\begin{document}

\date{Accepted 2010 February 17. Received 2010 February 16; in original form 2009 October 27}

\pagerange{\pageref{firstpage}--\pageref{lastpage}} \pubyear{2010}

\maketitle

\label{firstpage}

\begin{abstract}
A new harassment model is presented that models the complex, and dynamical tidal field of a Virgo like galaxy cluster. The model is applied to small, late-type dwarf disc galaxies (of substantially lower mass than in previous harassment simulations) as they infall into the cluster from the outskirts. These dwarf galaxies are only mildly affected by high speed tidal encounters with little or no observable consequences; typical stellar losses are $<10\%$, producing very low surface brightness streams ($\mu_B > 31$ mag arcsec$^{-2}$), and a factor of two drop in dynamical mass-to-light ratio. Final stellar discs remain disc-like, and dominated by rotation although often with tidally induced spiral structure. By means of Monte-Carlo simulations, the statistically likely influences of harassment on infalling dwarf galaxies are determined. The effects of harassment are found to be highly dependent on the orbit of the galaxy within the cluster, such that newly accreted dwarf galaxies typically suffer only mild harassment. Strong tidal encounters, that can morphologically transform discs into spheroidals, are rare occurring in $<15 \%$ of dwarf galaxy infalls for typical orbits of sub-structure within $\Lambda$CDM cluster mass halos. For orbits with small apocentric distances ($<$250 kpc), harassment is significantly stronger resulting in complete disruption or heavy mass loss ($>90 \%$ dark matter and $> 50 \%$ stellar), however, such orbits are expected to be highly improbable for newly infalling galaxies due to the deep potential well of the cluster.  
\end{abstract}


\section{Introduction}
Dwarf galaxies are easily the most common type of galaxy in the Universe, and yet their very origin remains a subject of much debate. They can be found in vast numbers in the cluster environment - our nearest large cluster, the Virgo cluster, contains over a thousand dwarfs. The understanding of the formation and evolution of cluster dwarfs is hampered by their intrinsic properties of low luminosity and low surface brightness causing them to be challenging objects to observe, and study. Intuitively, it might be expected that their low mass, and hence shallow potential wells, cause them to be extremely sensitive to their local environment. As such, cluster dwarfs aught to provide sensitive indicators to the effects of mechanisms occurring in the high density environment.

The Virgo cluster catalogue (\citealp{VCC}), continues to provide an invaluable tool for the study of dwarf galaxies in the Virgo cluster. Complete down to apparent magnitude $B_T>$18, and with a suprisingly accurate morphological classification that has, on the whole, stood the test of time, the Virgo cluster catalogue (VCC) clearly demonstrates that dwarf galaxies are far from a homogenous family of galaxies. Dwarf ellipticals are the most common type of cluster dwarf - typically containing no current star formation, and an old stellar population. Dwarf ellipticals can be additionally split into further categories, with some containing a compact, central stellar nucleus - the nucleated dwarf ellipticals, and dwarf spheroidals - generally considered to be low luminosity analogues of dwarf ellipticals. Not all the dwarfs are `red and dead' with many containing active star formation, with knotty HII regions clearly visible in the Las Campanas observatory photographic plates utilised to create the VCC. These star forming dwarfs are further split into sub-categories; dwarf irregulars containing atomic gas, and with evidence of rotation (\citealp{Zee2004}), and blue compact dwarfs whose gas and star formation is centrally concentrated and also show significant rotation (\citealp{Zee2001b}). A further classification is made for `transition-types', objects whose properties are somewhere mid-way between dwarf ellipticals and dwarf irregulars, and may form a possible evolutionary link between the two classes. The large variety in dwarf galaxy properties makes a single theory for their creation highly challenging.

With the release of SDSS data set, covering the Virgo region, further progress was made in the observation of cluster dwarfs. A series of papers (\citealp{Lisker2006},\cite{Lisker2006b},\cite{Lisker2007}) have provided a quantitive study of over 400 Virgo dwarf galaxies, suggesting that the VCC dwarf classifications can be further sub-categorised. The improved imaging quality, and additional colour information have shown that many dwarf ellipticals thought to be typically `red and dead', infact contain blue centres indicating recent or current central star-formation. Furthermore, dwarf ellipticals that appear smooth and featureless, present intricate spiral patterns and disks when an un-sharp mask technique is applied. \cite{Lisker2007} stress the importance of recognising these sub-categories and considering them separately when attempting to un-ravel dwarf galaxy origins, and conclude that multiple environmental mechanisms are required to explain the diverse properties of cluster dwarfs.

Current suggested dwarf galaxy formation and evolution scenarios are numerous and varied. The $\Lambda$CDM paradigm suggests that current day dwarf galaxies are the left over debris from hierarchical merging of haloes. While $\Lambda$CDM simulations predicts the presence of an extended dark matter halo about each dwarf galaxy, they make less detailed predictions about properties of the baryonic content of a galaxy. It is necessary to include physical mechanisms that influence the visible components of dwarf galaxies in order to bring the simulations into agreement with observations (\citealp{Bullock2000}). These physical mechanisms can be broadly split into two categories; global mechanisms that affect all dwarf galaxies regardless of their location, such as supernovae feedback (\citealp{Dekel1986}), and environmentally dependent mechanisms. One such environmental mechanism, known as harassment, is believed to be of importance in high density environments (\citealp*{Moore1998}, \citealp{Mastropietro2005}). Numerous high speed tidal encounters within a cluster act to morphologically transform medium mass, low surface brightness disc galaxies into smaller galaxies - cluster dwarfs. A second scenario, ram pressure stripping, acts on in-falling galaxies that are already dwarfs - late-type dwarfs - and that this can produce galaxies that resemble cluster dwarf ellipticals (\citealp{Zee2004}, \citealp{Boselli2008}). These two scenarios have one thing in common - a newly infalling disc galaxy is subjected to cluster environmental mechanisms causing transformation into an object resembling a cluster dwarf elliptical.  Indeed, radial velocity measurements of cluster dwarf ellipticals in the Virgo cluster do indicate that they are far from a virialised population within the cluster potential, bearing signatures of a recent in-fall (\citealp{Conselice2001}). Meanwhile dwarf galaxies obey their own morphology-density relationship; early -type dwarfs have been said to be the most strongly clustered of all galaxies, whereas gas-rich dwarfs are the most weakly clustered (\cite{Ferguson1994}). As cluster-centric distance increases, the ratio of dwarf ellipticals to dwarf irregulars decreases (\citealp{Sabatini2005}). These two observations combined may suggest that many cluster dwarf ellipticals have their origin in infalling late-type dwarfs.  Further evidence for galaxy transformation of discs is present in the form of hidden disc features, in some cases even including fine spiral arm structure (\citealp{Lisker2006b}, \citealp{Lisker2007}) in Virgo cluster dwarf ellipticals. It is as if the original discy nature of the galaxies has yet to be completely removed by the cluster environmental mechanisms. This is additionally supported by evidence of significant rotation in a sub-sample of the dwarf ellipticals in \cite{Zee2004b}. Remarkably the rotating dwarf ellipticals, and dwarf irregulars, both appear to obey the Tully-Fisher relation for bright late-type spirals. If we assume that a substantial fraction of present-day dwarf ellipticals are indeed accreted as dwarf discs into the cluster, the next question is whether it is harassment or ram pressure stripping that is the key environmental influence driving the tranformation. As ram pressure stripping has been shown to remove significant amounts of atomic hydrogen from {\it giant} spiral galaxies within the Virgo cluster (\citealp{Chung2008}, \citealp{Vollmer2002}), it should be of little surprise that ram pressure can strip small late-type dwarfs of their HI gas content, causing a cessation of star-formation, and evolving the dwarfs from blue to red colours, typical of cluster dwarf ellipticals (\citealp{Boselli2008}). However, some form of physical morphological transformation may additionally be required to transform dwarfs from thin discs to spheroidals. One suggestion is that both mechanisms are required - ram pressure halts the star formation, while harassment thickens up the disc.

In the literature (\citealp{Moore1998}), harassment is the effects of repeated and numerous long range tidal encounters in the potential well of the cluster. The tidal forces experienced by a galaxy within a cluster can be broadly split into two categories - tidal heating, and tidal shocking. The key difference between these categories is the time-scale on which the galaxy experiences the tidal force. For tidal shocking the time-scale is short - for example in a high speed tidal encounter between two cluster galaxies, a strong but short-lived tidal force may be experienced by the two galaxies. Tidal heating occurs when the time-scale is far longer - for example a cluster galaxy that passes close to the cluster centre spends a significant length of time under the influence of the deep potential well of the cluster, and may be entirely dismantled if the resulting tidal forces are in excess of the galaxy's self-gravity. Harassment is the combined effects of both tidal heating (from the tidal forces associated with the deep potential well of the cluster) and tidal shocking (from the tidal forces associated with the high speed galaxy-galaxy encounters) on a cluster galaxy.

The aim of this work is to study the effects of harassment on newly infalling, late-type dwarf irregulars. This is accomplished using a new, and fast algorithm to model a galaxy clusters tidal fields that produces similar results to far more complex models set in a cosmological context. In concordance with cosmological models, both dwarf galaxy models and the cluster tidal field utilise cuspy cold dark matter haloes.The dwarf galaxy models are significantly lower mass ($\sim$10$^9$ - $ 1 \times 10^{10}$ M$_\odot$) than previous studies of harassment on dwarf galaxies. \cite{Mastropietro2005} dwarf models were of mass $7 \times 10^{10}$ M$_\odot$ in comparison. Studying the effect of harassment on dwarfs in a lower mass regime is interesting for a number of reasons - obviously it extends harassment studies to cover a wider range of the parameter space. Additionally while larger dwarfs may be harassed into smaller objects resembling dwarf ellipticals (\citealp{Mastropietro2005}), the fate of smaller mass dwarfs is less clear. A weakly harassed low mass dwarf may be able to complete a morphological transformation into an object resembling a dwarf elliptical without significant mass-loss. Alternatively, a heavily harassed low mass dwarf may be entirely destroyed in the process. 

For these low mass dwarfs, the significance of tidal heating from the potential well of the cluster alone can be estimated using a Roche limit analysis. When the tidal forces of the cluster potential well exceeds the dwarf galaxy's own self-gravity, the dwarf will be entirely dismantled. For our standard dwarf model, and for the cluster potential utilised (see Section \ref{harassmodel}), this can be expected to occur at a radius of less than $\sim 50$ kpc within the cluster. Hence only dwarfs on extremely plunging orbits are significantly influenced by tidal heating. Hence for the orbits we consider (see Section \ref{simeffects}) whose peri-cluster distance is $\sim 200$ kpc, the effects of tidal heating alone are expected to be weak.

We can also analytically estimate the effects of tidal shocking from galaxy-galaxy encounters for our low-mass dwarf models. Although it seems intuitive, that the lowest mass galaxies might be expected to be the most significantly affected by tidal encounters, this is not necessarily the case. The strength of the effects of a high-speed encounter can be expected to depend on the dynamical time at the scale-radius, $r_s$ of the dwarf ($t_{dyn} \sim \left[(r_s^3/{G M(<r_s)}\right]^{\frac{1}{2}}$, where $G$ is the gravitational constant. When this dynamical time is long compared to the passage-time of a passing harasser galaxy, then the the dwarf can respond effectively to the encounter. It is interesting to note that in the dwarf models considered here, for material surrounding the disk of the galaxy, $t_{dyn}$ is of the same order as the passing time of a harasser galaxy. Despite the lowered mass, they are significantly more compact, and hence resilient to harassment. These properties are not peculiar to our models - in this mass regime, this is to be expected. 

Although tidal stripping and shocking are expected to be weak for the dwarfs we consider, strong tidal shocking between galaxies in combination with tidal heating can still produce significant effects. These effects are studied in detail on model dwarf galaxies in Section \ref{simeffects}. We attempt to quantify the likelihood of significant harassment in section \ref{Monty}, and test how the likelihood varies for a variety of orbits in Section \ref{orbitsection}.  \newline
\noindent 
The paper is organised as follows. In section 2, the numerical code utilised in all simulations is discussed, and in Section 3 the methods used to build model dwarf galaxies are shown. The harassment model is presented in section 4. In section 5, the influence of harassment on 11 in-falling dwarf galaxy models are shown. The statistical influences of harassment are found by means of a Monte-Carlo simulation in section 6, and tested for a variety of orbits.

\section[]{The code}
In this study we make use of `gf' (\citealp{Williams2001},\citealp{Williams1998}), which is a Treecode-SPH algorithm that operates primarily using the techniques described in \cite{Hernquist1989}. `gf' has been parallelised to operate simultaneously on multiple processors to decrease simulation run-times. While the Treecode allows for rapid calculation of gravitational accelerations, the SPH code allows us to include a HI gas component to our dwarf galaxy models. In all simulations, the gravitational softening length, $\epsilon$, is fixed for all particles at a value of 100 pc, in common with the harassment simulations of \cite{Mastropietro2005}. Gravitational accelerations are evaluated to quadropole order, using an opening angle $\theta_c=0.7$. A second order individual particle timestep scheme was utilised to improve efficiency following the methodology of \cite{Hernquist1989}. Each particle was assigned a time-step that is a power of two division of the simulation block timestep, with a minimum timestep of $\sim$0.5 yrs. Assignment of time-steps for collisionless particles is controlled by the criteria of \cite{Katz1991}, whereas SPH particle timesteps are assigned using the minimum of the gravitational time-step and the SPH Courant conditions with a Courant constant, $C$=0.1 (\citealp{Hernquist1989}). As discussed in \cite{Williams2004}, the kernel radius $h$ of each SPH paricle was allowed to vary such that at all times it maintains between 30 and 40 neighbours within 2$h$. In order to realistically simulate shocks within the SPH model, the artificial viscosity prescription of \cite{Gingold1983} is used with viscosity parameters $(\alpha,\beta)$ = (1,2). The equation of state for the gas component of the galaxies is isothermal with a bulk velocity dispersion of 7.5 km s$^{-1}$, based on the measured velocity dispersion of molecular clouds in the local interstellar medium (\citealp{Stark1989}). By choosing an isothermal equation of state, we are intrisically assuming that stellar feedback processes are balanced by radiative cooling producing a constant velocity dispersion. `gf' contains a simple star formation prescription, where each SPH particle converts it's mass into stellar mass at a rate controlled by a Schmidt law with a power of 1.5. When the global stellar mass formed in this manner reaches a specified mass, a specified number of new N-body star particles are formed. The new particles have their position chosen statistically such that recent star formation history of individual SPH particles is taken into account. For details of code testing, please refer to \cite{Williams1998}. 
 
\section{Building model dwarf galaxies}
Our dwarf galaxies models consist of 3 components; an NFW dark matter halo (\citealp*{Navarro1996}), an exponential disc of gas and one of stars. The methods used to form each component will be discussed in the following sections.
\subsection{The dark matter halo}
In the current standard model of galaxy formation galaxy mass is dominated by an unseen dark matter component, surrounding the disc. This provides the additional gravitational force required to explain the observed velocities while simultaneously producing their observed flat rotation curves. \cite{Navarro1996} suggests that the density profiles of dark haloes have a universal shape (the NFW profile), for haloes that range in mass from dwarf galaxies to galaxy clusters.

The NFW profile has the form:
\begin{equation}
\rho_{NFW}(r) = \frac{\rho_0}{(\frac{r}{r_s})(1+\frac{r}{r_s})^2}
\label{NFWdensprof}
\end{equation}
\noindent where $r_s$ is a characteristic radial scale-length. The profile is truncated at the virial radius, $r_{200}=r_s c$. Here $c$ is the concentration  parameter (\citealp{Lokas2001}). $c$ is found to have a range of values in cosmological simulation, however there is a general trend for higher values in less massive systems with some scatter - see figure 8, \cite{Navarro1996}. Typical values for cluster-mass objects is $c \sim$4, whereas dwarf galaxies can have $c \sim 20$. For our standard dwarf galaxy model we choose a total mass of $10^{10}$M$_\odot$, and $c$=20. Positions and velocities are assigned the dark matter particles using the distribution function described in \cite{Widrow2000}.

Dark matter haloes produced in this manner are tested by evolving the initial conditions using the numerical simulation code, and the evolution of the density profile is observed. Tests show that after 0.5 Gyrs, transient effects have settled, and the halo remains stable. Due to the abrupt truncation of the density distribution at $r_{200}$, $\sim 2 \%$ of halo mass can expand outwards at the truncation, increasing it's radii by $\sim 25 \%$ but this causes negligible effect on the density distribution surrounding the disc.

\subsection{The galaxy discs}
Within the centre of the NFW halo is a rotating stellar and gas disc. The stellar and gas disc of typical spiral galaxies are observed to be exponential, albeit with roughly a factor of two difference between the gas and stellar scalelength (\citealp{Boselli2006}):

\begin{equation}
\label{expdisc}
\Sigma(r) = \Sigma_0 exp (r/r_s)
\end{equation}

\noindent
where $\Sigma$ is the surface density, $\Sigma_0$ is central surface density, $r$ is radius within the disc, and $r_s$ is the scale-length of the disc.

In order to choose a reasonable scale-length for the stellar component of the disc, we follow the recipe of \cite{Mo1998} for producing discs within the $\Lambda$CDM paradigm. This recipe has been shown to reproduce the Tully-Fisher relationship's slope and scatter, as well as the observed scatter in the size-rotation velocity plane. Having first chosen a halo mass, the disc mass $M_d$ is assumed to be a fixed fraction $m_d$ of the halo mass. Additionally, the angular momentum of the disc is assumed to be a fixed fraction $j_d$ of the halo's angular momentum. \cite{Mo1998} find $m_d \sim j_d$ and $m_d \leq$0.05 to reproduce observations, and we choose $m_d$=0.05 for all galaxy models. The scale-length of the disc is then fully defined by the properties of it's host halo:

\begin{equation}
r_s = \frac{\lambda G M_{200}^{3/2}}{2 V_{200} \mid E \mid ^{1/2}} \left(\frac{j_d}{m_d}\right)
\end{equation}
where $M_{200}$, $V_{200}$,and $E$ are total mass, circular velocity, and energy of the halo, and $\lambda$ is the spin parameter of the halo. We choose $\lambda$=0.05 for all disc models presented in this paper unless otherwise stated (the mean $\lambda$ in the  \cite{Mo1998} probability function) . For our standard galaxy model, this produces an $r_s$=0.86 Kpc. Knowing this, we can solve for $\Sigma(r)$ in Equation \ref{expdisc} using $\Sigma_0 = M_d/2 \pi r_s^2$. The disc scalelength for the gas disc is then simply the stellar disc scalength multiplied by a factor of two. In order to set up the positions of stellar and gas particles, we integrate over $\Sigma(r)$ in concentric rings using a standard integral such that

\begin{equation}
\label{discmwirad}
M(r') = \int_0^{r'} \Sigma(r) 2 \pi r dr = 2 \pi \Sigma_0 {r_s}^2 \left[ \left(\frac{-r'}{r_s}-1\right)e^{-r'/r_s}+1\right]
\end{equation}

\noindent As with the dark matter halo, we can solve for $r'(M)$ numerically, then by producing a random number between 0 and 1 for the normalised M, we are provided with the radius of each particle. These particles are initially laid down in a single plane. Real discs have a finite thickness, but we cannot solve for this until we have defined the velocity dispersion throughout the disc. This is necessary in order to ensure that the disc is Toomre stable (\citealp{Toomre1964}). The Toomre stability criterion is defined as

\begin{equation}
\label{Toomreeqn}
Q \equiv \frac{\kappa\sigma_r}{3.36 G \Sigma} > 1
\end{equation}

\noindent
Here $\Sigma$ is the surface density, $\sigma_r$ is the radial velocity dispersion, and $\kappa$ is the epicyclic frequency defined, using the epicyclic approximation (\citealp{SpringelWhite1999}). Next we use $\sigma_\phi^2 = \frac{\sigma_r^2}{\gamma^2}$ where $\gamma^2 \equiv \frac{4}{\kappa^2 R} \frac{d\Phi}{dR}$ (also \citealp{SpringelWhite1999}), and $\phi_z = 0.6 \cdot \phi_r$ (\cite{Shlosman1993}). This completely defines the minimum values of the necessary velocity dispersions throughout the disc. In practice, $Q>1.5$ is required throughout the stellar disc to ensure stability. The gas disc has an intrinsic velocity dispersion, due to it's isothermal nature, that automatically satisfies the Toomre criteria at all radii. Once more following \cite{SpringelWhite1999}, we now use $z_d = \frac{\sigma_r^2}{\pi G \Sigma}$ for the vertical scale height of the disc, and distribute the particles vertically out of the disc following the Spitzer's isothermal sheet solution

\begin{equation}
\rho(R,z) = \frac{\Sigma(r)}{2z_d} {\rm sech}\,^2(z/z_d)
\end{equation}.

Finally, the circular velocities of disc particles are calculated. The value of the potential is calculated in thin spherical shells for the combined dark matter, stars and gas in a one-off N-body calculation, including the effects of gravitational softening. This is necessary as the discs own self-gravity can influence the radial orbits of disc particles close to the centre of the halo. The gradient of the potential is then used to calculate the circular velocity at each radius. As a result inner disc particles have raised circular velocities beyond that of the circular velocity of the halo alone at that radius.

In all standard dwarf galaxy model simulations, a gas-rich dwarf irregular is simulated, containing $4 \times 10^8$ M$_\odot$ of gas and $1 \times 10^8$ M$_\odot$ of stars. Such gas rich systems can be found in Virgo cluster dwarf galaxy observations (\citealp{Boselli2006}). However the paper's conclusions are additionally tested using a pure stellar galaxy disc model to verify the significance of the gas-to-stellar ratio on the influences of harassment - see Section \ref{diffmodel}.

To ensure long term stability, the complete three-component galaxy models are evolved in isolation for 2.5 Gyrs. Transient effects are found to fully settle in $<$1 Gyr and only evolved models are introduced into the cluster harassment models.

A summary of the standard dwarf galaxy models properties is provided in Table \ref{galprops}.

\begin{table}
\centering
\begin{tabular}{|c|c|c|}
\hline
 Component & Parameter & Parameter value\\
\hline
Halo& Total mass & 10$^{10}$ M$_\odot$ \\
& Concentration $c$ & 20\\
& Virial Radius $r_{200}$& 44 kpc\\
& Particle number & 200000\\
Total disc & Mass & 5 $\times 10^8$ M$_\odot$\\
& Particle number & 50000 \\
Stellar disc & disc fraction & 0.2 \\
& Particle number & 10000 \\
& Scalelength $r_d$ & 0.87 kpc \\
Gas disc & disc fraction & 0.8 \\
& Particle number & 40000 \\
& Scalelength $r_d$ & 1.74 kpc \\
\hline
\end{tabular}
\caption{Summary of key parameters of standard dwarf galaxy model}
\label{galprops}
\end{table}
 
\section{The harassment model}
\label{harassmodel}
The potential well from the gravitational field of a large galaxy cluster, such as Virgo, applies a significant force on nearby galaxies. The local group itself is thought to be infalling towards the cluster at $\sim$200 km s$^{-1}$ as a result of the Virgo over-density (\citealp{Tammann1985}). However, it is the {\it tidal} force of the cluster that will strip or deform an in-falling galaxy. Observationally, clusters have typical mass-to-light ratios of several hundred (\citealp{Sheldon2009}). As a result, the combined mass of individual galaxies, even including their dark matter haloes, is not sufficient to total the mass of a cluster. 

As a result, we model the tidal field of a Virgo-like cluster using a two component model; the first is a dynamical tidal field associated with individual `harasser' galaxies, and the second is a static tidal field to represent the rest of the mass of the cluster that is not associated with galaxies - referred to as the background cluster potential herein. The potential field of both the harassers and the background is represented using the analytical form for an NFW density distribution (\cite{Lokas2001})

\begin{equation}
\label{potNFW}
\Psi = -g_cGM_{200}\frac{ln(1+(r/r_s))}{r}
\end{equation}
\noindent where $g_c=1/[{ln(1+c)-c/(1+c)}]$.

\begin{figure}
\centering%
\epsfxsize=5.9cm 
\begin{rotate}{
\begin{rotate}{
\begin{rotate}{\epsfbox{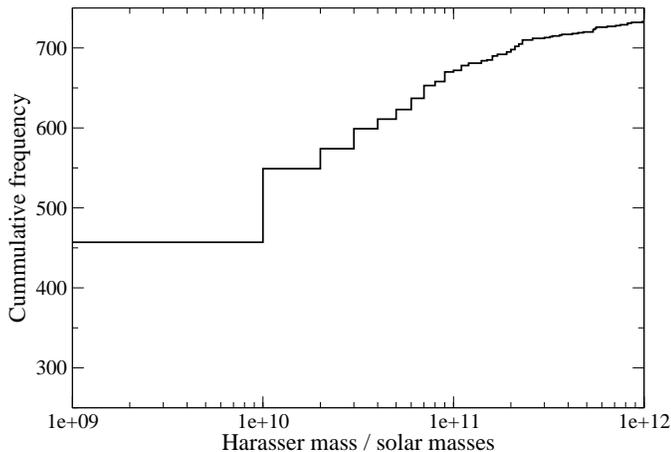}}
\end{rotate}}
\end{rotate}}
\end{rotate}
\caption{Cummulative mass distribution of harasser galaxies, binned in $10^{10} M_\odot$ width bins}
\label{galmassdistrib}
\end{figure}

Modelling the total cluster potential as two components has an additional advantage - it is easy to separate and study the effects of the background cluster potential from the additional effects of high-speed galaxy-galaxy encounters. The background cluster potential is fully defined once its virial mass and concentration have been chosen. We choose $c$=4, and $M_{200}$=1.6$\times$10$^{14}$M$_\odot$. This choice of cluster mass is in agreement with the model Virgo cluster potential of \cite{Vollmer2001}, and with galaxy line-of-sight velocity measurements in \cite{VCC} ($\sim 10^{14}$M$_\odot$). Now each harasser galaxy has it's own individual NFW potential field. To make this field dynamical, a one-off N-body simulation of a cluster-mass halo is conducted, and the time-dependent coordinates of a fraction of the particles are logged. This log now provides the position-evolution of the centre of each of the harassing haloes, where each individual galaxy potential field will be super-imposed. Once more we require a mass and concentration for each harasser's potential field to be defined. For mass, we use the Schector function (\citealp{Schechter1976} with parameters provided in \cite{Sandage1985}, and assume a constant mass-to-light ratio, i.e.

\begin{equation}
M_{tot} = \frac{\phi_\star}{L_\star} \left(\frac{M}{L}\right)_K \int_{L_{min}}^{\infty} \left( \frac{L}{L_\star} \right)^{\alpha + 1} \exp \left( -\frac{L}{L_\star} \right) dL
\label{massfunct}
\end{equation}

\noindent
where $\alpha=-1.25$, and $(M/L)_K=20$. The fitting parameters provided in \cite{Sandage1985} are based on the Virgo cluster catalogue (\citealp{VCC}), and are therefore complete down to a minimum galaxy luminosity of $\sim 2.5 \times 10^7 L_\odot$ (assuming a distance modulus of 31.0 for Virgo). We choose to resolve all harasser galaxies down to one-tenth the mass of our infalling dwarf model ($10^9 M_\odot$). Completing the integral in Equation \ref{massfunct} produces a total mass in galaxies that is 14.5$\%$ of the total cluster mass. This is in reasonable agreement with the $\sim 10 \%$ percentage of a cluster's mass locked up in sub-structure, found in $\Lambda$CDM simulations - see \cite{Gill2004}. This process produces 733 harassing galaxies in total. The cummulative distribution of galaxy masses produced can be seen in Figure \ref{galmassdistrib}. The vast majority of harasser galaxies are dwarfs, with $\sim 65 \%$ of the total number of harassers having a mass less than the standard dwarf galaxy model, while only $\sim 9 \%$ of the harassers have a mass greater than 10 times that of the model dwarf. The maximum mass galaxies produced are $\sim 10^{12} M_\odot$. The value of the mass-to-light ratio chosen is consistent, if not an upper limit for galaxies in this mass-range (see \citealp{Gilmore2007}). The concentration $c$ of each halo is chosen to follow the trend of higher concentration for lower mass objects as found in $\Lambda$CDM simulations. This is achieved by a fit to concentration values with mass found in \cite{Navarro1996} producing $c = -3 \log (M_{halo}(M_\odot)) + 52$. This fit produces $c \sim 15$ for a $6 \times 10^{11}$ M$_\odot$ mass halo, and $c \sim 20$ for a $1 \times 10^{10}$ M$_\odot$ mass halo.

It should be noted that the use of analytical potentials for the harassing galaxies tidal fields has advantages and disadvantages over previous harassment models. \cite{Mastropietro2005} and \cite{Moore1999} utilise full $\Lambda$CDM simulations of the dark matter in cluster-mass objects, while \cite{Moore1998} utilises softened point masses to represent the tidal fields of harassers. These previous methods have the advantage that the harassing galaxies can interact and react to the tidal potential of the harassed galaxy model, whereas analytical potentials on fixed paths of motion do not. This essentially violates Newton's third law - the dwarf can respond to the cluster potential field, but the cluster potential field cannot respond to the dwarf. However, primarily we are interested in studying the effects of the harassers on the model galaxy and not vice versa. As will be shown in Section \ref{locencs}, encounters velocities are typically very high, and as a result the tidal forces in a high speed encounter are brief, and strongly peaked. The ability for the harasser to alter it's trajectory in response to the dwarf models tidal field is inconsequential to the brief, strong tidal forces that the model is subjected to at these velocities. Typically the dwarf model suffers a slight deviation in it's orbit as the result of a strong tidal encounter, and the same might be expected for the harasser galaxy. However this will have a negligible effect on the strength, and shape of the peaked tidal forces due to the rapid passing time, and high velocity of the encounter. Additionally, the use of analytical potentials has advantages - the tidal forces have effectively infinitely small spatial resolution. Only the self-gravity of the galaxy model's own particles have a limited spatial resolution due to softened gravity. Analytical potentials are highly computationally cheap, allowing a large number of simulations to be conducted rapidly as is a necessity for harassment simulations due to their highly stochastic behaviour (see Section \ref{Monty}).

\section{Effects of harassment on model dwarf galaxies}
\label{simeffects}
The harassment model is utilised for 8 infalls (Runs 1-8) of the standard dwarf galaxy model. Initially, the dwarf galaxy models are positioned at the outskirts of the cluster, and randomly orientated on a shell at the virial radius ($r_{apo}$=1.1 Mpc), such that each dwarf galaxy is presented with a different set of tidal encounters as it infalls. They are each given an azimuthal velocity of 250 km s$^{-1}$. If all harassing galaxies are removed, this produces a plunging orbit with a pericenter of $\sim$ 200 kpc and a peak velocity of $\sim$ 1600 km s$^{-1}$. This choice of orbit is consistent with the range of orbits utilised for infalling galaxies in \cite{Vollmer2002}. The orbit is chosen to model a dwarf galaxy infalling into the cluster {\it for the first time} at the current epoch. Models are evolved for 2.5 Gyrs providing time for the infalling galaxy to make one-pass of the cluster centre on this orbit. In order to complete a second pass, the model would need to be evolved for $>$ 6 Gyrs in which case the assumption of constant cluster mass becomes invalid. Hence the model is most applicable to a first infall into the cluster environment. The effects of less plunging orbits are simulated and discussed in Section \ref{Monty}.

\subsection{Tidal histories}
\begin{figure}
\centering%
\includegraphics[scale=0.34,angle=-90]{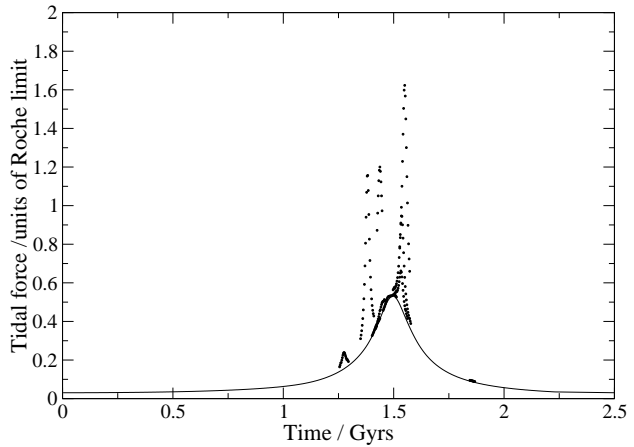}
\caption{A tidal history for Run 4. Tidal forces are measured in units of the isolated dwarf galaxy model's Roche limit at the edge of it's stellar disc. The Gaussian-like curve represents the contribution from the back ground cluster potential, and points are the contribution from individual galaxy-galaxy encounters.}
\label{run4_tidehist}
\end{figure}

As we know both the position-evolution and the properties of all harasser galaxies, we can measure the exact tidal force, from the background cluster potential and from harasser galaxies, felt by an infalling galaxy at any instant. All tidal force measurements are presented in units of the Roche limit of the isolated dwarf galaxy model at the edge of its stellar disc - a discussion of the use of Roche limits is provided in the following section. In Figure \ref{run4_tidehist}, the tidal evolution of Run 4 is presented as a function of time. We refer to these figures as {\it `tidal histories'}, and Run 4 can be considered a typical infall for reasons that will be further discussed in Section \ref{Monty}. Clearly the background cluster potential alone doesn't produce tidal forces beyond the dwarf galaxy's Roche limit, even at the peri-center of the orbit. As discussed and quantified in the Introduction, this would require a far more deeply plunging orbit. However, it is the sum of both the background and the individual encounters that can produce tidal forces exceeding the Roche limit. 

\subsection{Mass loss}
\begin{figure}
\centering
\epsfxsize=9.0cm \epsfbox{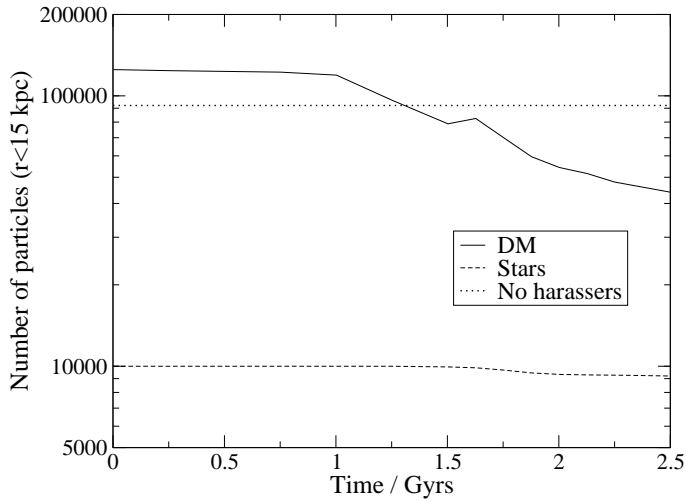}
\caption{Number evolution of dark matter particles (solid), and star particles (dashed), within a sphere of radius 15 kpc centred on the stellar disc for Run 4. The dotted line shows the final number of dark matter particles at the end of a simulation with the background cluster potential alone, i.e. no harassers included}
\label{run4_npartevol}
\end{figure}

Although the net tidal force rises above the Roche limit, it does so only briefly and hence the complete dismantling of the dwarf does not occur. This form of tidal force is therefore in the `tidal-shocking category' (see Introduction for further discussion). The use of tidal forces expressed in units of the Roche limit of the dwarf is directly relevant for tidal heating, but not so for tidal shocking.  As a result, it should be clearly noted that the presenting of all tidal forces in units of the Roche limit serves merely as a more intuitive and comprehendable scale or yard-stick for the strength of the high speed tidal encounters. In figure \ref{run4_npartevol}, the number evolution of dark matter, and stellar particles within a sphere of radius 15 kpc centered on the stellar disc is presented. Clearly the stellar disc is far from completely dismantled, and loses $< 10 \%$ of it's original mass. The dark matter halo is affected far more significantly, losing $\sim 60 \%$ of it's original mass. To test that this quantification of dark matter loss is robust to increased resolution, we repeat the infall with the gravitational resolution improved by a factor of 10 (the softening length, $\epsilon$ is reduced by a factor of 10 to 10 pc). We also improve the dark matter mass resolution by a factor of 2.5 by conducting the high resolution simulation with a dark matter halo consisting of 500,000 dark matter particles. The high resolution models are found to have negligible differences in dark matter and stellar mass losses from the lower resolution studies. To confirm how much of the dark matter loss is from the background potential alone, and how much is lost because of the inclusion of the harassing galaxies, the simulation is repeated without the harassing galaxies. In this simulation, the galaxy models loses only $\sim 25 \%$ of its dark matter - shown as dotted line in Figure \ref{run4_npartevol}. Therefore galaxy-galaxy interactions are responsible for more than half of the dark matter losses. As discussed in \cite{Moore1998}, the reason that the stars are so less significantly affected is due to their circular, centralised orbits whereas the eccentric orbits of dark matter particles render them far more susceptible to tidal disruption. The net result of enhanced dark matter losses over stellar losses is a decrease in dynamical mass-to-light ratio by roughly a factor of 2. With currently available data sets, it would be a significant observational challenge to detect a systematic difference in dynamical mass-to-light ratio of isolated dwarfs in comparison to cluster dwarfs at this level, due to the inherent observational and theoretical difficulties in obtaining these ratios. Such a test would provide a significant test of our understanding of dark matter. As discussed in \cite{Moore1998}, it is difficult to see how cluster galaxies can avoid losing a sizeable fraction of their dark matter, within the $\Lambda$CDM paradigm.

\subsection{Little morphological transformation}
The final harassed stellar disc {\it does not} appear to be significantly transformed from the original disc. The discs remains flattened and dominated by rotation. The isolated dwarf galaxy has a peak circular velocity  V$_{peak} \sim $80 km s$^{-1}$, and a velocity dispersion $\sigma$ out of the plane of the disc $\sim$ 10 km s$^{-1}$. After harassment, $V_{peak}$ falls slightly, and $\sigma$ is raised to 14 km s$^{-1}$. This has a net effect of reducing the ($V_{peak}/\sigma$) ratio by less than half to $\sim 4.6$, therefore remaining significantly dominated by rotation. 

\begin{figure}
\centering%
\includegraphics[scale=0.9]{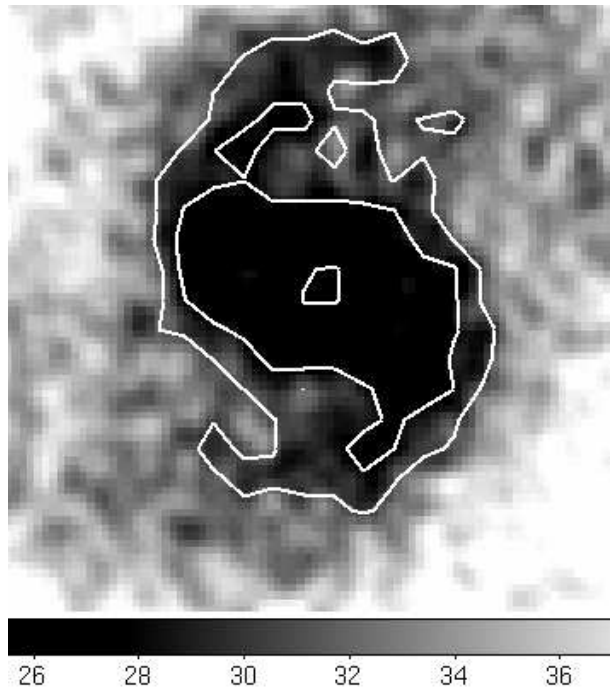}
\caption{B-band surface brightness plot of face-on stellar disc in Run 3 after 2.5 Gyrs of harassment. The grey-scale colour bar indicates surface brightness in units of mag arcsec$^{-2}$. The box size is 5 $\times$ 5 kpc. White isochrones are displayed with values of 25, 28, and 31 mag arcsec$^{-2}$ (from galaxy centre to edge of disc)}
\label{spiralinduced}
\end{figure}

The post-harassment stellar discs, which show induced spiral structure and bars, were initially smooth and featureless - see Figure \ref{spiralinduced}. It is interesting to note that in \cite{Lisker2007}, spiral structure in dwarf ellipticals is observed, and presented as evidence for the origin of dwarf ellipticals in disc galaxies. However, in these simulations such features originate {\it because} of high speed tidal encounters. The stars that are stripped are arranged along very low surface brightness streams, that move along similar orbits to that of the galaxy though the cluster. Note that in Figure \ref{spiralinduced}, the extended low surface brightness streams are not visible even at surface brightness limits of $\mu_B > 31 $ mag arcsec$^{-2}$. To form these images, a stellar mass-to-light ratio of $\sim 6$ is assumed following \cite{Mastropietro2005}. These surface brightnesses are well beneath the surface brightness limits of current optical Virgo galaxy surveys. For example, the INT wide-field survey has $\mu_B \sim 26$ mag arcsec$^{-2}$ (\citealp{Davies2005}).

Although we have concentrated on one typical harassment infall simulation, 6 out of 8 of the dwarf galaxy infalls are affected in a very similar manner to the results presented above. In Table \ref{8simeffects}, the effects of harassment in terms of mass loss and change in ($V_{peak}/\sigma$) is summarised for all 8 simulations. The exception is Run 5 and Run 6 for which, in both cases, the galaxy model has at least one encounter of $\sim 4$ times the Roche limit or greater. To better understand the frequency of such encounters we utilise  Monte-Carlo simulations (see Section \ref{Monty}).

\begin{table}
\centering
\begin{tabular}{|c|c|c|c|c|c}
\hline
 Run & Tides & DM & Star & ($V_{peak}/\sigma$) & (M/L)$_{dyn}$\\
\hline
1& 2.4,2.5,1.4& 30 & 82 & 45 & 37 \\
2& 2.5 & 59 & 100 & 59 & 59\\
3& 1.1,1.8,2.3 & 65 & 94 & 69 & 70\\
4& 1.1,1.2,1.6 & 41 & 93 & 55 & 44\\
5& 1.8,1.7,5.6 & 5 & 22 & 30 & 22\\
6& 3.7,1.9,1.8 & 8& 34 & 44 & 23 \\
7& 1.6,1.1,1.2 & 33 & 91 & 57 & 36 \\
8& 2.3 & 47 & 98 & 58 & 48 \\
\hline
\end{tabular}
\caption{Summary of 8 dwarf galaxy infalls; {\it column 1} is run description, {\it column 2} is summary of all tidal encounters with tidal strength greater than the Roche limit of the isolated dwarf galaxy, {\it column 3} and {\it column 4} are percentage of dark matter and stellar mass at the end of the simulation, found within a 15 kpc radius sphere, {\it column 5} and {\it column 6} is final ($V_{peak}/\sigma$), and final (M/L)$_{dyn}$ as a percentage of the orignal}
\label{8simeffects}
\end{table}

\subsection{Effects on gas and star formation}
\label{gaseffects}
In Figure \ref{gasspiralinduced}, the effects of high speed tidal encounters on the gas component of the infalling dwarf can be seen. Prior to the infall ({\it left panel}) the gas disc is a smooth exponential disc. Tidal encounters can induce spiral structure, which is stronger than is seen in the stellar disc due to the dissipative nature of the gas ({\it central panel}). The deformation of the disc from a circular shape causes shocking and barring of the gas disc, resulting in enhanced star formation within the spiral arms. Loss of angular momentum along the spiral arms can additionally cause radial inflow of the gas into the central regions of the dwarf irregular. However the spiral structure is only temporarily induced and, once the driving mechanism has stopped, the disc becomes smooth once more ({\it right panel}), leaving only a central knot of gas. The short-lived enhancement of gas densities within spiral arms and the central region results in a total galaxy star formation rate that is bursty in nature (Figure {\ref{burstysfrs}). The star formation rates of the dwarf irregular models decline steadily as they convert their gas reservoir into stars. However tidal encounters can cause both brief lowering (when the gas disc is stretched and lowered in density), and raising (when the gas disc is compressed) of the star formation rates. It should be noted that the simplicity by which star formation and associated feedback processes have been modelled here, provides only qualitative predictions for star formation rates. For example, a prescription for super-nova feedback, and gas cooling is not specifically included. Instead, it is assumed that energy input from stellar feedback is balanced by radiative cooling to produced an isothermal, single phase gas representing the diffuse, atomic gas content of the galaxy. The artificial viscosity of the gas component does enable it to shock, and this is important for spiral-arm formation, and radial gas inflow towards the galaxy centre. We do not include a prescription for gas-ionisation from background UV photons that may be important in galaxies of this mass (\cite{Mayer2006}). Hence, the star-formation rates seen in these simulations can only be considered qualitive at best, based purely on the density of the gas in the disk. 

However, it is likely that the missing gas physics is of little consequence as infalling dwarf galaxies will be most significantly influenced by ram pressure stripping in the cluster environment. The effects of ram pressure stripping will be investigated in detail in a subsequent paper (\citealp{Smith2010a}). For now, we restrict our analysis to the use of a simple analytical prediction for the efficiency of ram pressure stripping (\citealp{GunnGott1972}). Despite its simplicity, the Gunn and Gott formalism has been proven to provide reasonable first order predictions for the gas truncation radius in agreement with numerous full, numerical simulations of ram pressure stripping (\citealp{Abadi1999}, \citealp{Vollmer2002}, \citealp{Mayer2006}, \citealp{Jachym2007}, \citealp{Roediger2007}). We use the intra-cluster medium radial density profile of \cite{Vollmer2002}, for the varying infall velocity of our infalling dwarfs, to predict the radius within the cluster when the dwarf's gas disc will be completely stripped. This method predicts that the dwarf will be completely stripped at a cluster radius of $\sim 350$ kpc, {\it before} it completes it's first pass of the cluster centre. Thus, it is unlikely that the majority of infalling dwarfs will still contain appreciable amounts of gas before they experience harassment. The typical location within the cluster that an infalling dwarf experiences harassment is discussed further in Section \ref{locencs}. Additionally this excludes the possibility that a central nucleus could form from the radial inflow of gas seen in the harassment simulations (although it should be noted that nuclei tend to be observed more frequently in brighter dwarf galaxies (\citealp{Lisker2007}) that may be able to maintain a truncated HI disk after ram pressure stripping, and hence could allow formation of nuclei in this manner).

\begin{figure*}
\centering%
\includegraphics[scale=0.5]{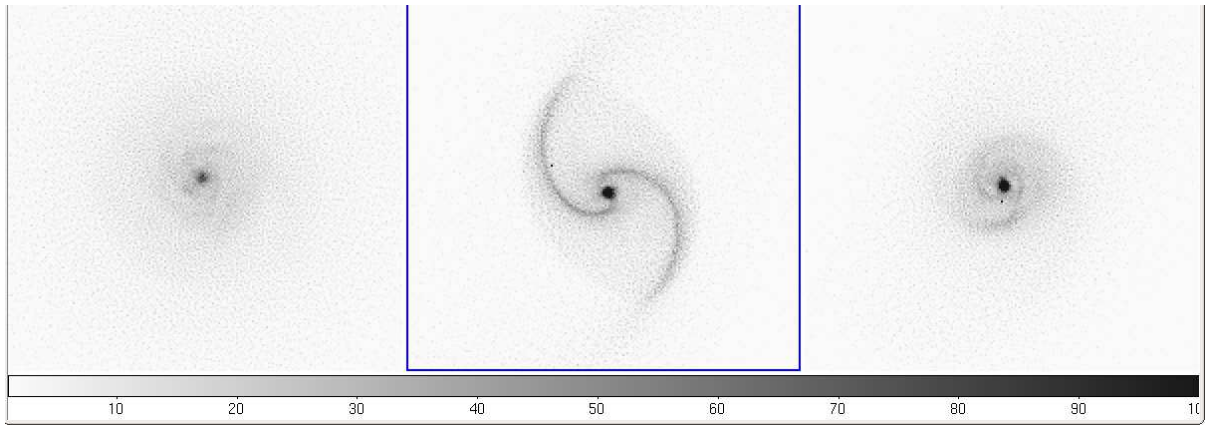}
\caption{Column density plot of face-on gas disc in Run 4;  ({\it left}) before harassment, ({\it centre}) post tidal encounter, ({\it right}) at the end of the simulation. The grey-scale colour bar indicates column density in units of M$_\odot$ pc$^{-2}$. The box size is 10 $\times$ 10 kpc}
\label{gasspiralinduced}
\end{figure*}

\subsection{Dependency on disc properties}
\label{diffmodel}
Our harassment model is ideal for testing how a galaxys response to harassment depends on specific properties such as gas fraction, surface brightness and mass. As will be shown in Section \ref{Monty}, a tidal history such as that of Run 4 can be considered a statistically typical infall. Now different galaxy models can be made to infall along the same trajectory, and be subjected to an identical tidal history. Using this method we test how different dwarf galaxy's respond to harassment; (i) Model A contains no gas, (ii) Model B is a more low surface brightness disc, and (iii) Model C is $\sim 6$ times less massive.

\begin{figure}
\centering
\epsfxsize=9.0cm \epsfbox{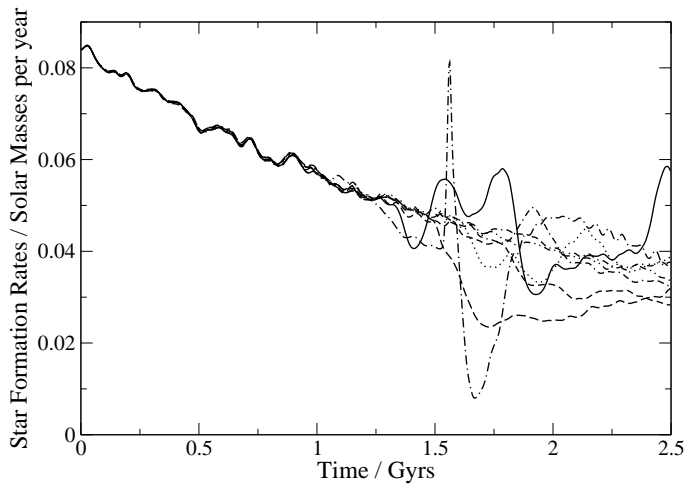}
\caption{Bursty star formation rates of the 8 infalling dwarf irregular models. Star formation rates are shown in M$_\odot$ yr$^{-1}$}
\label{burstysfrs}
\end{figure}

\subsubsection{Model parameters}
Model A is similar to the standard dwarf model, except it's disc mass is purely stellar. Each gas particle is replaced with an identical mass stellar particle. Model B is also similar except that we utilise a spin parameter $\lambda=0.1$. This is double the value of the standard dwarf galaxy model and forces the exponential disc scalelength of the stellar and gas disc to also double, to $r_d=1.72$ kpc (stars) and 3.48 kpc (gas). As a result, model B is $25 \%$ lower in surface brightness at all radii within the disc than in Model A. Model C has a dark matter halo of $1.6 \times 10^{9}$ M$_\odot$. As it's dark matter halo is less massive, it's concentration is increased slightly to $c=24$. Following the \cite{Mo1998} recipe for this halo, produces a stellar disc with an exponential scale length $r_d=0.43$ kpc. Once more, the scalelength of the gas disc is double that of the stellar disc.

\subsubsection{Response to typical harassment}
In Table \ref{diffgalresp} the key effects of harassment are summarised for ease of comparison. Particles are counted within a 15 kpc radius, to calculate dark matter and stellar losses in Model A. A sphere of double this radius was used for Model B, to match the fact that it's stellar disc had doubled in diameter. A sphere of only 7.5 kpc was used for Model C to account for it's smaller stellar disc.

\begin{table}
\centering
\begin{tabular}{|c|c|c|c|c|c}
\hline
 Run & DM & Star & ($V_{peak}/\sigma$) & (M/L)$_{dyn}$\\
\hline
Standard& 41 & 93 & 55 & 44 \\
Model A (star only)& 42 & 91 & 61 & 46\\
Model B (LSB)& 41 & 75 & 34 & 55\\
Model C (low mass)& 37 & 97 & 73 & 38\\
\hline
\end{tabular}
\caption{Summary of effects of typical harassment on varying dwarf galaxy models; {\it column 1} is run description, {\it column 2} and {\it column 3} are percentage of dark matter and stellar mass at the end of the simulation,{\it column 4} and {\it column 5} is final ($V_{peak}/\sigma$), and final (M/L)$_{dyn}$ as a percentage of the orignal}
\label{diffgalresp}
\end{table}

Firstly, comparing the standard model to Model A, there appears to be very little change in the effects of harassment in terms of mass loss, mass-to-light ratio, or stellar dynamics. A visual inspection of the stellar disc also reveals little difference, with mild enhancement of spiral structure in both models. The inclusion of a gas component has made little difference. This suggests thats the effects of harassment will not change significantly if ram-pressure stripping had removed the gas from the galaxy prior to harassment (see Section \ref{gaseffects} for further discussion). However, a deeper study of the combined effects of harassment and ram pressure stripping is deferred to a future paper (\citealp{Smith2010a}).

The same cannot be said for the low surface brightness disc of Model B. The halo has suffered similar losses as the standard model - this is to be expected as their haloes are identical and suffering the same tidal interactions. But the stellar disc suffered additional $\sim 15 \%$ losses. The ($V_{peak}/\sigma$) ratio is also reduced by an additional $10\%$ from the standard model. The simplest explanation is that a larger stellar disc represents a larger target for harassing galaxies. In addition, the self gravity of the disc is reduced by a factor of $25 \%$. This is in agreement with the simulations of \cite{Moore1999} where a low surface brightness giant spiral suffers considerably more harassment than a high surface brightness spiral. The surface brightness (or disc scale-length for exponential discs) appears to be an additional parameter controlling a galaxys sensitivity to harassment, as it controls the cross-sectional area of a galaxy to harassers.

Finally, we compare the low mass dwarf (model C) to the standard model. Despite model C having considerably less mass, its dark matter halo has not suffered substantially greater mass-loss. Furthermore, it's stellar disc losses are a smaller mass fraction than in the standard model. The ($V_{peak}/\sigma$) ratio suffers only a minor reduction in comparison to the other models. This further supports the conclusion that the cross-sectional area of a galaxy is a strong parameter controlling the response of a galaxy to harassment.

\section{A Monte-Carlo simulation of harassment}
\label{Monty}

Column 2 of Table \ref{8simeffects} summarises the tidal histories of each infalling dwarf galaxy model. There is a large range in strength and frequency of tidal encounters. This brings into question just how commonly strong encounters occur, or in fact just how typical our `typical' simulation really is. To attempt to answer this question, and to gain additional insight into the principles of harassment, we conduct a 1000 galaxy infall Monte-Carlo simulation.

Clearly, conducting 1000 full N-body/SPH simulations of infalling dwarf galaxy models would be extremely time-consuming. Instead 1000 N-body particles with mass equal to the dwarf galaxy (and with the same assumed Roche limit) are allowed to infall into the cluster model. Initially each particle is randomly positioned on a spherical surface of radius 1.1 Mpc. They are given an initial velocity of 250 km s$^{-1}$ and the direction of the velocity vector on the spherical surface is randomised. Hence the initial orbital parameters ($r_{apo}$ and $v_{azi}$) of each particle's orbit is identical to that used in the full numerical simulations. However, now the range of tidal histories that an infalling dwarf galaxy encounters is sampled to a far greater degree of completeness than is possible with only 8 galaxy model infalls. The tidal history of each particle is then recorded allowing for a statistical analysis, and by comparison with the tidal histories of the full simulations, conclusions can be drawn. 

\subsection{Frequency of strong encounters}

\begin{figure}
\centering%
\epsfxsize=5.9cm 
\begin{rotate}{
\begin{rotate}{
\begin{rotate}{\epsfbox{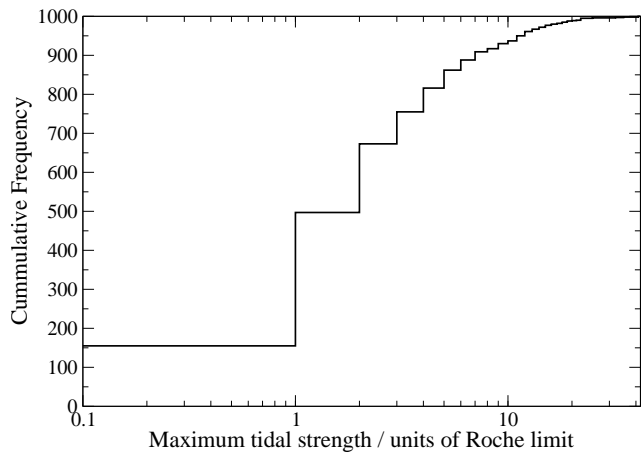}}
\end{rotate}}
\end{rotate}}
\end{rotate}
\caption{Cummulative frequency plot of peak strength tidal encounters experienced by infalling dwarf galaxies on plunging orbits}
\label{cummfreqplunge}
\end{figure}

To understand, statistically, what the typical tidal history of an infalling dwarf galaxy appears like, we log the peak tidal force that each infalling Monte-Carlo particle is subjected to throughout its orbit. In Figure \ref{cummfreqplunge}, a cummulative frequency plot of the peak strength encounter is shown. It can be seen that half of the infalling dwarfs never experience a peak tidal force greater than twice the Roche limit of the dwarf model. In fact peak tidal forces in the range of 1-2 Roche limits are the most common experienced by the infalling galaxies. In the full numerical simulations, both Run 4 and Run 7 have this type of tidal history, so these can be considered as typical infalls. It should be noted that even tidal histories that include peak tidal forces as large as 3 Roche limits (occurring in two-thirds of infalls) do not show significantly stronger effects of harassment (see Table \ref{8simeffects}). Significant tidal encounters that cause strong transformation or stellar losses occur for tidal forces that exceed $\sim 4$ Roche limits in the full numerical simulations. These occur in less than $25 \%$ of the infalls and cannot be considered statistically typical. Hence mild harassment can be considered the norm for newly accreted dwarf galaxies, at least for the orbits considered so far (see Section \ref{orbitsection} for the dependency of this conclusion on orbital parameters).

\begin{figure}
\centering%
\epsfxsize=6.0cm 
\begin{rotate}{
\begin{rotate}{
\begin{rotate}{\epsfbox{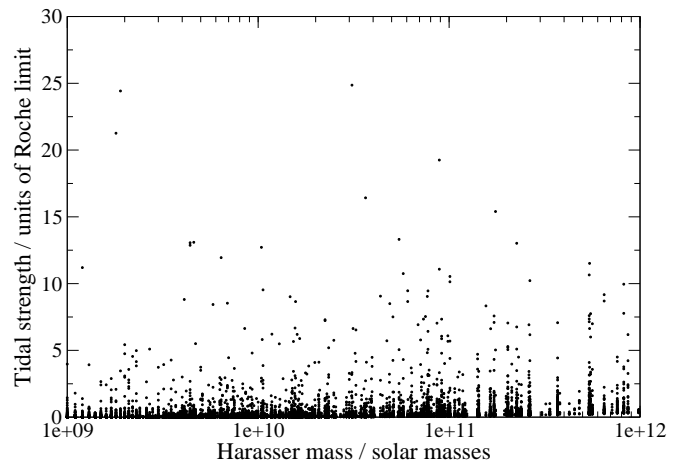}}
\end{rotate}}
\end{rotate}}
\end{rotate}
\caption{Peak tidal strength of encounters is plotted against mass of the harassing galaxy. There is no indication of stronger tidal encounters occurring at the high end of the galaxy mass function}
\label{strengthvsmass}
\end{figure}

\subsection{What drives strong tidal encounters?}
We are also in a position to answer the question of whether strong tidal forces are dominated by encounters with the rarer, more massive harassing galaxies. In Figure \ref{strengthvsmass}, the strength of the encounter versus the mass of the harassing galaxy is plotted. There is a large degree of scatter in the plot and little trend for stronger tidal encounters with increased mass of the harasser. Strong tides are not predominantly caused by encounters with the rarer, more massive cluster galaxies. A far stronger trend is seen when we instead plot tidal strength against the minimum separation between a harassing galaxy and the infalling dwarf - see Figure \ref{strengthvsrad}. This clearly indicates that strong tides are in fact driven by close encounters with harasser galaxies. This is contrary to statements in \cite{Moore1996} suggesting that the bulk of the evolution is caused by encounters with galaxies brighter than $\sim L_\star$. In fact the strongest encounters seen may not be well modelled by the simulations in this paper, as the proximity of the galaxy-galaxy encounter is such that a collision of the baryonic contents is likely. If the colliding galaxies both contain a dissipative gas component, then there is the possibility for a dissipative (wet) merger. Fortunately, this is not a common event. $<6 \%$ of all encounters occur at separations of less than 15 kpc. The dwarf galaxy model initially has a gas disc that extends to $\sim 15$ kpc, hence passes closer than this, between similar galaxies, are likely to cause a collision rather than a tidal encounter. The maximum tidal strength encountered at a separation greater than 15 kpc is 5 times the Roche limit of the dwarf galaxy. Thus infalling galaxies who experience tidal forces as strong as this (e.g. Run 5) can still be considered to be well modelled by the simulations in this paper. However, it is not unreasonable for us to assume that the infalling dwarf galaxies are ram pressure stripped of their gas content before the collisions occur. In this case, the collisionless stellar component of the dwarf galaxy will only be affected by gravitational tides, as modelled in these simulations, and the results of even stronger tides would still remain reasonable. This has interesting implications for the possibility of collisionless (dry) mergers. Newly infalling galaxies are not captured by the tidal potentials of the harassers galaxies, but merely deviate from their original trajectory.

\begin{figure}
\centering
\epsfxsize=9.0cm \epsfbox{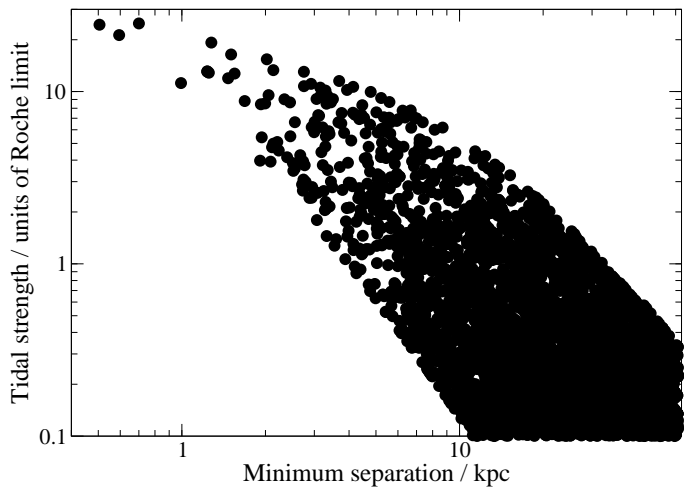}
\caption{Tidal strength of encounters is plotted against minimum separation between the dwarf galaxy and the harassing galaxy. There is a strong trend for increasing strength of tidal encounters with decreasing separation.}
\label{strengthvsrad}
\end{figure}

\subsection{Location and speed of encounters}
\label{locencs}
While ram pressure stripping is believed to be only significant in the inner regions of the cluster, how common are high speed tidal encounters in the outskirts of the cluster? In this model, harassment is found to be similarly preferential to the cluster centres. Statistically, the average radius of an encounter is $\sim 200 \pm 150$ kpc within the cluster. Only $\sim 7 \%$ occur at radii beyond 500 kpc (beyond the core radius as defined in \cite{Sabatini2005}). Hence, the effects of harassment are very unlikely to be observed in the outskirts of the cluster, unless the dwarf galaxy has already made a pass of the cluster centre (i.e. it is a member of the back-splash (\citealp{Gill2005}) galaxy population).

The average velocity of an encounter is very high with a large standard deviation: $\sim$ 2000 $\pm$ 600 km s$^{-1}$. Hence the short-lived strong tides seen in Figure \ref{run4_tidehist} are virtually the standard and consequently galaxy-galaxy capture events are very rare . Chance low speed encounters are vanishingly small, with $< 4 \%$ occurring at velocities of $<500$ km s$^{-1}$. At first glance, these encounters appear to occur at very high velocity, considering the total velocity dispersion of the cluster is $\sim 1000 $ km s$^{-1}$. However, the elliptical orbits of the harasser galaxies causes the velocity dispersion in the central region of the cluster to be higher
and this is where the majority of encounters occur. In addition, the plunging orbit of the infalling dwarf is far from the orbit of a virialised galaxy population raising encounter velocities further.
\subsection{Strong sensitivity to orbit}
\label{orbitsection}
\subsubsection{A less plunging orbit}
The orbit utilised in the full numerical simulations and in the Monte-Carlo simulation described above, is not un-realistic but does involve a plunging orbit. We repeat the Monte-Carlo simulation but this time with a less plunging orbit. This allows us to test how strongly the tidal histories of infalling dwarf galaxies depend on their orbit through the cluster.

For this we use the average orbital parameters for sub-structure in cluster-mass objects of a $\Lambda$CDM simulation - see \cite{Gill2004}. Despite a wide variety of cluster properties, the orbits of the satellite populations were found to be similar with an average ellipticity $e=0.61$, and an average pericenter distance $r_{peri}=0.35 r_{vir}$ (with minimal scatter in both quantities). Here ellipticity is defined as $e=1-r_{peri}/r_{apo}$ and $r_{vir}$ is the virial radius of the cluster. For the cluster model presented in this paper, this fixes $r_{peri}=385$ kpc, and $r_{apo}=987$ kpc. Hence the orbit is less eccentric, and does not plunge so deeply into the cluster centre. To produce this orbit, all of the Monte-Carlo particles representing galaxies are positioned at the apocenter of the orbit, and given a 450 km s$^{-1}$ azimuthal velocity. This produces a peak velocity at pericluster distance of $\sim 1250$ km s$^{-1}$.

The result is significantly less strong harassment. In Figure \ref{cummfreqtypical} a cummulative frequency plot of peak tidal forces encountered by a dwarf galaxy following a typical orbit is shown. Now $70\%$ of galaxy infalls never experience a tidal force greater than two times the Roche limit, compared to $50 \%$ for the previous more plunging orbit. Such a tidal history typically resulted in only minor stellar losses, and minor disc heating in the full numerical simulations - see Table \ref{8simeffects}. Still Run 4 and Run 7 represent the most statistically common results of harassment. Strong tidal encounters that cause significant stripping and morphological transformation (at tidal forces $\sim 4$ times the dwarf galaxies Roche limit or more) occur in less than $15 \%$ of the infalls, compared to $25 \%$ for the previous plunging orbits utilised. Clearly, the influences of harassment are strongly dependent on the orbital parameters of the orbiting galaxies, and this can be simply understood - the deeper the orbit plunges in to the cluster, the more harassing galaxies are encountered. However, there must be an additional dependency on the time spent in the cluster core also. Hence it might be expected that the strongest harassment will occur for much less eccentric orbits, that are close to circular, and remain near to the cluster center throughout their orbits.

\begin{figure}
\centering
\epsfxsize=9.0cm 
\epsfbox{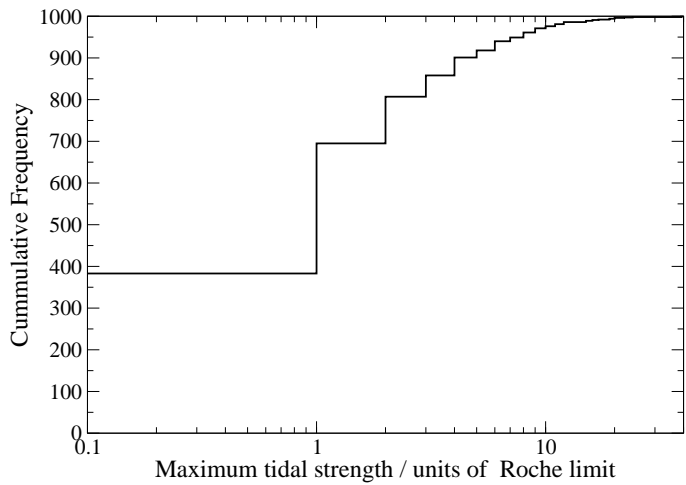}
\caption{Cummulative frequency plot of peak strength tidal encounters experienced by infalling dwarf galaxies on typical orbits for substructure within $\Lambda$CDM simulations of cluster mass dark matter haloes}
\label{cummfreqtypical}
\end{figure}

This does indeed seem to be the case for the dwarf-harassment simulations in \cite{Mastropietro2005}. Galaxies that have large apocentric distances ($\sim r_{vir}$) all only suffer very mild harassment with similar dark matter and stellar losses to the dwarf galaxies in these simulations ($\sim 10 \%$ stellar and $\sim 60-70 \%$ dark matter lost). Although the harassment model used in this study involves analytical calculation of tidal forces, it produces similar results to a full $\Lambda$CDM simulation of a cluster, although a direct comparison is not possible due to the significantly lower mass of the dwarf models in this study. Meanwhile in \cite{Mastropietro2005}, almost all orbits whose apocentre is $\sim 0.1$ r$_{vir}$, and hence never leave the dense cluster core suffer significant mass loss ($\sim 50 \%$ and greater stellar loss, and $\sim 95 \%$ dark matter loss). A low eccentricity, small apo-centric orbit is additionally more likely to make a number of repeat passes of the cluster core. If a galaxy makes a second pass, the chances of a strong encounter should double if the cluster mass and number of harassers has not changed significantly over this time. The strongly harassed dwarf galaxies in \cite{Mastropietro2005} complete numerous orbits over the length of the simulation. Similar conclusions are also drawn in \cite{Gill2004}. However, this form of orbit is {\it highly unlikely} to occur for galaxies that are {\it newly accreted} into the cluster at the current epoch. The cluster now has a larger mass, and the galaxy's orbit must have high eccentricity to pass close to the cluster core where the effects of harassment are strong. This results in a long period orbit with a large apo-centre, where it is unlikely that the galaxy will make a second pass in $>6$ Gyrs (for the NFW background potential used in this study). Dwarfs that make numerous repeat cluster center passes must have formed close to the cluster core, or at least infallen when the cluster was significantly less massive. Such dwarfs could form their own sub-class of dwarf, as discussed in the `Discussion and Summary' section.

\subsubsection{A small apo-centric orbit}
To test the influence of harassment on galaxies that are not newly infalling into the cluster, we repeat the Monte-Carlo simulations only this time we place each Monte-Carlo particle closer to the centre of the cluster, at a radius of 200 kpc. Each is provided with a purely azimuthal velocity of 660 km s$^{-1}$ producing an orbit that, in the absence of harasser galaxies, is virtually circular (zero eccentricity). A cummulative frequency plot of the peak tidal force experienced by each of the Monte-Carlo particles is presented in Figure \ref{montecarlo_central}.

\begin{figure}
\centering
\epsfxsize=9.0cm 
\epsfbox{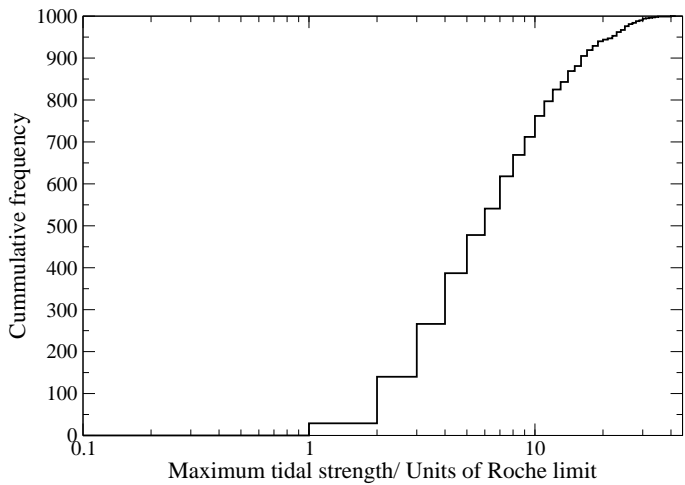}
\caption{Cummulative frequency plot of peak strength tidal encounters experienced by infalling dwarf galaxies initially on a circular orbit, close to the cluster centre at radius 200 kpc}
\label{montecarlo_central}
\end{figure}

Clearly, strong tidal encounters occur significantly more often for such a small apo-centric orbit. There are now {\it no} Monte-carlo particles that experience a peak tidal force of less than one Roche limit over 2.5 Gyrs of evolution. In fact $>$70$\%$ of all particles experience at least one tidal encounter of force greater than 4 times the Roche limit of the dwarf, which has been shown to cause significant dark matter and stellar losses. Hence for this type of orbit, strong harassment is the norm. To test the effects of such tidal encounters on small dwarf galaxies, we place the standard model dwarf galaxy at a 200 kpc radius and provide it with the same 660 km s$^{-1}$ aximuthal velocity. We then evolve the dwarf galaxy for 2.5 Gyrs under the influences of harassment. We repeat this test 4 times, with a different velocity vector each time (although all have a purely azimuthal velocity of 660 km s$^{-1}$) so as the 4 models each under go different tidal histories. The results can be seen in Table \ref{centraltable}.

\begin{table}
\centering
\begin{tabular}{|c|c|c|c|c|c|c}
\hline
 Run & Apo-Peri & DM & Star & ($V_{peak}/\sigma$) & (M/L)$_{dyn}$\\
\hline
Run 1& 220-100 & 12 & 64 & 38 & 19 \\
Run 2& 200-50 & 1 & 3 & - & -\\
Run 3& 240-70 & 1 & 8 & - & -\\
Run 4& 240-120 & 28 & 85 & 55 & 33\\
\hline
\end{tabular}
\caption{Summary of effects of a small apocentric orbit  (r$<$250 kpc) harassment on the standard dwarf galaxy model; {\it column 1} is run description, {\it column 2} is apocentre and pericentre of the orbit, {\it column 3} and {\it column 4} are percentage of dark matter and stellar mass at the end of the simulation,{\it column 5} and {\it column 6} is final ($V_{peak}/\sigma$), and final (M/L)$_{dyn}$ as a percentage of the orignal}
\label{centraltable}
\end{table}

Run 1 suffers strong harassment, in a similar manner to the rare strong encounters of the more plunging orbits, (see Run 6 of Table \ref{8simeffects}) with $\sim 90 \%$ dark matter losses, and a substantial decrease in ($V_{peak}/\sigma$) and (M/L)$_{dyn}$. Run 2 and Run 3 are both destroyed by tidal encounters, making a measurement of their final ($V_{peak}/\sigma$) and (M/L)$_{dyn}$ impossible. The Monte-carlo tests demonstrate that disruption by the cluster background potential alone is a rare ocurrence. $<$1$\%$ of Monte-carlo particles experience a tidal force above 90$\%$ of the models Roche limit from the cluster background potential alone. Run 2 and 3 are dismantled by repeated strong tidal encounters and pass the cluster centre several times over the period of their orbit. Run 4 suffers mild harassment - similar to the typical harassment of the plunging orbits (e.g. Run 1 and Run 7 of Table \ref{8simeffects}). It can be considered as one of the $<$30$\%$ of the Monte-carlo particles that avoids tidal encounters of greater than four times the Roche limit, hence is fairly atypical of harassment in this regime. 

In general, the dwarfs are typically destroyed or suffer significant dark matter and stellar losses when undergoing small apocentric orbits. The final stellar disks of the two surviving galaxies show significant heating, and appear to have increased in vertical scale-height substantially. It is estimated that, if viewed edge-on, the ratio of their short-axis to long-axis would be $\sim 0.8$ and $\sim 0.5$ in Run 1 and Run 4 respectively - hence Run 1 has become close to spherical in it's stellar distribution.

Once more, a direct comparison between the results of \cite{Mastropietro2005} is not possible due to the differing mass-regime of the dwarf galaxy models, but substantial dark matter and stellar losses of equivalent magnitudes ($\sim$90 $\%$ and $\sim$50 $\%$ respectively, and greater) are seen. For comparison Gal 1,2,4,7-12 of Table 1 in \cite{Mastropietro2005} have approximately similar orbital characteristics and mass-loss.  However, one difference is that none of Mastropietro's dwarfs were considered completely destroyed by the end of their simulation. The lower mass dwarfs are more susceptible to disruption by the combined action of the background cluster potential (which is continuously stronger than in the plunging orbit case), and numerous strong tidal encounters. In this situation, dark matter that is liberated by a strong encounter, is quickly swept away by the stronger tidal field of the background cluster potential. This brings into question the fate of the dark matter and stars from such disrupted galaxies. One possibility is a fraction of the dark matter and stars is absorbed into the central elliptical of the clusters, while any stars that are not absorbed in this manner contribute to the low surface brightness background observed in the cluster (\citealp{Mihos2005}). Indeed, the filaments and streams of stars observed at low surface brightnesses between the giant central elliptical galaxies in Virgo, may have their origin in the merging of heavily disrupted harassed dwarfs with the giant ellipticals.

\section{Discussion and Summary}
As discussed in \cite{Lisker2007}, it is important that we recognise the various sub-classes of dwarf galaxy if we are to understand their origin. If we split the cluster dwarfs into nucleated and non-nucleated dwarfs, we find significant variations between the two populations. In order to recover the intrinsic, three dimensional shape of dwarf galaxies, flattening distributions combined with the surface brightness test are utilised (\citealp{Binggeli1995}, \citealp{Lisker2007}). Comparisons of {\it non-nucleated} dwarf ellipticals, with dwarf irregulars and late-type spirals show no statistically significant difference in their flattening distributions as measured by a standard Kolmogorov-Smirnov test. However \cite{Binggeli1995} find that {\it nucleated} dwarfs maybe as oblate as classical ellipticals, and \cite{Lisker2007} find they are more akin to E3 or E4 type giant ellipticals. Further evidence for a different evolutionary history is reflected in their spatial distribution. Nucleated dwarfs are found to be concentrated towards the centre of the Virgo and Fornax cluster like the giant E and S0 galaxies. Meanwhile non-nucleated dwarfs are distributed like the late-type spirals and irregulars, suggesting they are far from virialised within the cluster (\citealp{Ferguson1989}). A similar conclusion is drawn in \cite{Lisker2007}, who uses radial velocity profiles to suggest that the nucleated dwarf ellipticals are indeed virialised, whereas the non-nucleated dwarfs still bear the signatures of a recent infall.

This accumulation of evidence could suggest that nucleated dwarfs formed in situ within the cluster, whereas the non-nucleated population have been accreted from the cluster outskirts. The research presented in this paper suggests that harassment does not provide a significant mechanism to convert newly infalling late-type dwarfs into dwarf ellipticals. Ram pressure stripping of late-type dwarfs may provide a far more effective mechanism to form the non-nucleated dwarfs (\citealp{Boselli2008}) and will be studied in further detail in \cite{Smith2010a} in combination with the harassment model presented in this paper. 

However, it is less clear how ram pressure stripping could have formed the nuclei found in the nucleated dwarf population. The simulations of \cite{Mastropietro2005} demonstrate that harassment may be far more effective on a more centrally concentrated, and virialised dwarf population that formed in situ in the cluster. The early cluster environment would be expected to have lower intra-cluster medium densities, and so ram pressure stripping would be far less efficient than it is today. If their inter-stellar medium is not stripped, then tidal encounters could efficiently drive gas into their centres (\citealp{Moore1998}), and could facilitate the formation of central nuclei. Perhaps these galaxies were the progenitors of current day nucleated dwarf ellipticals whose oblate shapes may reflect the stronger harassment they have incurred.

\subsection{Key results}

\begin{enumerate}
  \item In general ($>75 \%$ of infalls), dwarf galaxy models that infall from the outskirts of the cluster suffer only mild harassment. Typical stellar losses are $\sim 10 \%$, while typical dark matter losses are $\sim 60 \%$, resulting in a reduction in dynamical mass-to-light ratio by a factor $\sim 2$. Typical ($V_{peak}/\sigma$) ratios are also reduced by approximately the same factor, but final stellar discs remain rotationally dominated, and appear as thickened discs, and often with tidally induced spiral structure.
  \item In rarer cases ($< 25 \%$ of infalls), dwarf galaxy models infalling from the cluster outskirts suffer strong harassment, losing $\sim 70 \%$ of their stars and $\sim 90 \%$ of their dark matter.
  \item A Monte-Carlo simulation reveals that strong harassment occurs in $< 25 \%$ of infalls for the deeply plunging orbit, and $<15 \%$ of infalls for a more typical orbit of substructure within a $\Lambda$CDM cluster-mass halo. The effects of harassment are strongly dependent on orbital parameters, as this dictates time spent in the cluster centre and the number-density of harassers encountered. 
\item For a low eccentricity, small apo-centric orbit, strong harassment is typical, occurring in $>70 \%$ of galaxies on such orbits over 2.5 Gyrs. This typically results in strong mass loss ($>90\%$ dark matter, $>50 \%$ stars) or complete disruption of these low mass dwarf models.
  \item Loss of stellar mass is caused by impulsive heating of the stellar disc by short-lived tidal interactions. Stars that are stripped from the disc by harassment form very low surface brightness streams ($\mu_B > 31$ mag arcsec$^{-2}$)
  \item The Monte-Carlo simulation reveals that average tidal encounters occur at high velocities ($\sim 2000$ km s$^{-1}$), with chance low speed encounters being very rare ($< 4 \%$ of encounters have a relative velocity $<500$ km s$^{-1}$).
  \item For a plunging infall, galaxy-galaxy encounters are centrally concentrated, typically occurring at a radius $\sim 1/5$th the cluster virial radius.  Encounters in the outer cluster are rare ($\sim 7 \%$ occur at a cluster radius $>500$ kpc). Therefore the effects of harassment are unlikely to be observed in the outskirts of the cluster for newly accreted galaxies - with the exception of the back-splash population that has already completed one near pass of the cluster centre.
  \item Strong tidal interactions are not primarily caused by encounters with massive harasser galaxies, but are driven by near-pass encounters with more typical cluster galaxies.
  \item Spiral structure such as those observed in \cite{Lisker2006} can be induced by tidal encounters, and therefore is not necessarily an indication that that original galaxy contained the same spiral structure. 
       
\end{enumerate}

\section{Conclusions}
Harassment of small late-type dwarf galaxies {\it that have recently been accreted into the cluster} is not an effective mechanism for forming the thickened discs of dwarf ellipticals. Despite their low mass, they are additionally very concentrated in mass, and hence are robust against the effects of harassment. In previous studies (\citealp{Moore1998}), harassment is the effect of repeated long-range fast encounters. For dwarfs in the low-mass but more concentrated regime studied here, harassment is only signicant in repeated, or even individual, close-range high speed encounters where tidal forces are the strongest, and these are rare for objects infalling into the cluster at the current epoch. Currently it would be a challenge to observe the typical effects of harassment on dwarfs that have recently infallen into the cluster (very low surface brightness extragalactic features, mild disc thickening, and a $\sim 50 \%$ reduction in dynamical mass-to-light ratios). However harassment appears far more effective on dwarf galaxies that have formed in situ within the cluster as it formed, possibly forming nucleated cluster dwarfs. Alternatively harassment of larger, low surface brightness spirals into dwarf ellipticals in the manner suggested by \cite{Moore1998} is another possibility. However, the simulations of \cite{Moore1999} demonstrate the harassment is not effective on all disc galaxies - while a giant low surface brightness disc is effectively tranformed, a high surface brightness giant like our own galaxy is only mildly affected. Therefore if cluster dwarf ellipticals have their origins in larger disc galaxies, a very large population of specifically low surface brightness field disc galaxies is required, in order to form the vast numbers of dwarf ellipticals seen in clusters today.
The formation of dwarf ellipticals from dwarf irregulars via ram pressure stripping is another convincing possibility as discussed in \cite{Boselli2008}. Ram-pressure stripping need only terminate their star formation to produce objects that appear very much like dwarf ellipticals today (\citealp{Zee2004}).

\section*{Acknowledgments}

Greatest thanks to Dr M. Fellhauer for a critical reading of the original version of the paper, and to Dr L. Cortese for numerous helpful discussions and support. Thanks also to Rhys Taylor whose input was vital in developing simulation visualisation tools. Finally, thanks to the School of Physics and Astronomy, Cardiff University who enabled our usage of the Coma Cluster Super Computer for all simulation presented in this work. This research was enabled and funded by an STFC studentship.
\bibliography{bibfile}

\bsp

\label{lastpage}

\end{document}